\documentclass[%
    reprint,
    amsmath,
    amssymb,
    aps,
    pra,
    longbibliography,
    showkeys,
    superscriptaddress,
]{revtex4-2}

\usepackage{graphicx}
\usepackage{dcolumn}
\usepackage{hyperref} 

\usepackage{bm}
\usepackage{xcolor} 
\usepackage[per-mode=symbol]{siunitx} 

\DeclareSIUnit\angstrom{\text{\AA}}         
\DeclareSIUnit\elementarycharge{\text{e}}   
\DeclareSIUnit\atm{\text{atm}}              
\DeclareSIUnit\amu{\text{amu}}              

\hypersetup{colorlinks=true, 
    linkcolor={blue},
    citecolor={blue}, 
    urlcolor={blue}
}

\begin{document}

\title{Chiral Surface Phonons}

\author{Mike Pols}
\email{mike.pols@mat.ethz.ch}
\affiliation{%
    Materials Theory, ETH Zurich, CH-8093 Z\"urich, Switzerland
}%

\author{Nicola A. Spaldin}
\email{nicola.spaldin@mat.ethz.ch}
\affiliation{%
    Materials Theory, ETH Zurich, CH-8093 Z\"urich, Switzerland
}%

\date{\today}

\begin{abstract}

We use symmetry arguments combined with density functional theory to demonstrate that surfaces of centrosymmetric crystalline materials host surface phonons that are chiral. As model systems, we study slabs of highly symmetric AB rocksalt compounds, and find surface-localized phonons whose atomic displacements exhibit chiral motion. We further show that these chiral surface phonons generate in-plane magnetic moments at the surface. Our results indicate that chiral phonons can emerge in all crystalline materials as a result of reduced symmetry at surfaces or interfaces. These findings establish surfaces as a previously overlooked source of chiral phonons and their associated magnetic moments, which could play a role in surface-sensitive measurements.

\end{abstract}


\maketitle

\section*{Introduction}

Quantized vibrations of the atoms in a crystal, or phonons, play an important role in thermal transport~\cite{cahillLatticeVibrationsHeat1988}, structural stability~\cite{cowleyAcousticPhononInstabilities1976}, and light-matter interactions of materials~\cite{kampfrathResonantNonresonantControl2013}. Phonons with a well-defined sense of rotational motion, commonly referred to as \textit{chiral phonons}~\cite{juraschekChiralPhonons2025}, have attracted considerable attention in recent years. In these vibrations, atoms follow circular or elliptical trajectories, resulting in a finite phonon angular momentum $\mathbf{J}$~\cite{zhangChiralPhononsHighSymmetry2015, zhuObservationChiralPhonons2018} and helicity $\hat{\mathbf{q}} \cdot \mathbf{J}$~\cite{juraschekChiralPhonons2025}. Chiral phonons break improper rotational symmetries and are found in chiral crystals and more generally in non-centrosymmetric materials~\cite{cohClassificationMaterialsPhonon2023, yangSymmetryguidedCatalogueChiral2026}. Their angular momentum can cause a phonon magnetic moment~\cite{juraschekDynamicalMultiferroicity2017, juraschekOrbitalMagneticMoments2019}, resulting in a range of novel phenomena, including the phonon Zeeman effect~\cite{juraschekDynamicalMultiferroicity2017, chengLargeEffectivePhonon2020, baydinMagneticControlSoft2022}, a phononic analogue of the Einstein--de Haas effect~\cite{zhangAngularMomentumPhonons2014, dornesUltrafastEinsteinHaas2019, zhangMeasurementPhononAngular2025}, and magnetization switching via the phonon Barnett effect~\cite{daviesPhononicSwitchingMagnetization2024, basiniTerahertzElectricfielddrivenDynamical2024}.

Symmetry lowering at surfaces and interfaces is known to be a source of a variety of electric and magnetic phenomena, including surface magnetization in antiferromagnets~\cite{weberSurfaceMagnetizationAntiferromagnets2024, weberLocalMagnetoelectricEffects2025} and surface multiferroicity~\cite{bhowalEmergentSurfaceMultiferroicity2025}. For phonons, it is well established that surfaces give rise to surface-localized phonon modes~\cite{chenSurfacePseudosurfaceModes1971, allenStudiesVibrationalSurface1971a, allenStudiesVibrationalSurface1971b, chenStudiesVibrationalSurface1977, wallisSurfaceVibrationalProperties1985, kressSurfacePhonons1991}, which can exhibit elliptical ionic motion, as a consequence of the reduced symmetry at the surface~\cite{allenStudiesVibrationalSurface1971a}. However, the connection between such surface phonon modes and concepts of phonon angular momentum and chirality, as well as their potential magnetic effects, remains unexplored.

Here, we use density functional theory (DFT) to investigate the possible chiral nature of phonons at crystal surfaces. We focus on AB rocksalt compounds, i.e. NaCl, RbF, and CsH, as prototypical high-symmetry model systems. Our calculations reveal chiral surface phonons, with handed ionic motion that generates in-plane magnetic moments. A general symmetry analysis further shows that chiral surface phonons are symmetry-allowed at all crystal surfaces due to broken $\mathcal{P T}$ symmetry, with $\mathcal{P}$ and $\mathcal{T}$ denoting inversion and time-reversal symmetry, respectively. Consistent with this symmetry argument, we also find chiral surface phonons at the surfaces of other materials, including copper and graphite. Our results establish chiral surface phonons as a general property of crystal surfaces, and show that their magnitude depends on the orientation and microscopic atomic structure of the surface.

\begin{figure}[htbp]
    \includegraphics{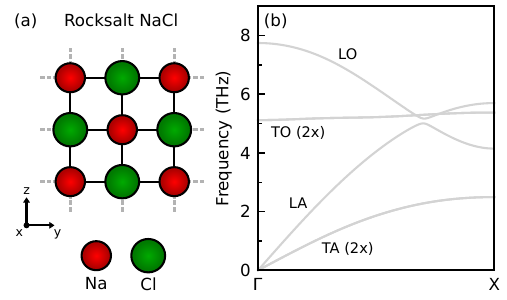}
    \caption{Phonons of bulk rocksalt NaCl. (a) Crystal structure of rocksalt NaCl. The red and green spheres represent Na$^{+}$ and Cl$^{-}$ ions, respectively. Dashed gray lines indicate bonding through periodic boundaries. (b) Calculated bulk phonon dispersion with modes labeled as acoustic (A) or optical (O), and further classified as transverse (T) or longitudinal (L). Doubly degenerate phonon modes are indicated with $2\times$.}
    \label{fig-1:bulk_phonons}
\end{figure}

\begin{figure*}[htbp]
    \includegraphics{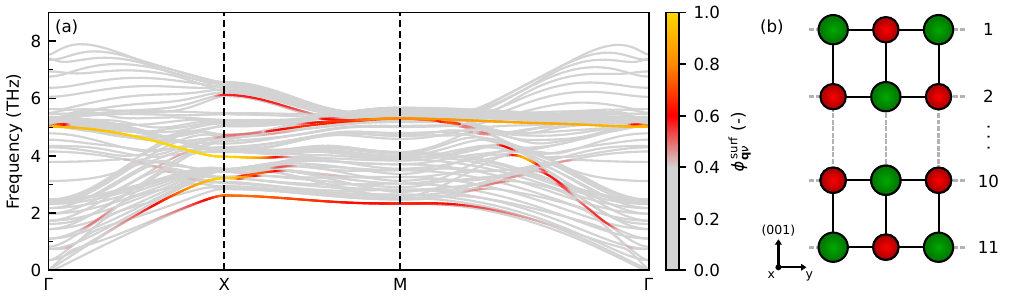}
    \caption{Surface phonons in rocksalt NaCl. (a) Slab phonon dispersion, with the phonon bands colored with the surface localization $\phi^{\mathrm{surf}}_{\mathbf{q} \nu}$. A value of $\phi_{\mathrm{surf}} = 1$ indicates complete localization on the outermost layers of the slab. The $\mathrm{\Gamma - X}$ path for bulk rocksalt NaCl is equivalent to the $\mathrm{\Gamma - M}$ path in the slab. (b) 11-layer (001)-oriented NaCl slab. The top and bottom surfaces are layers 1 and 11, respectively. Dashed gray lines indicate bonding through periodic boundaries or into the slab.}
    \label{fig-2:surface_phonons}
\end{figure*}

\section*{Computational methods}

We perform calculations with DFT as implemented in the Vienna Ab-initio Simulation Package (\texttt{VASP})~\cite{kresseInitioMoleculardynamicsSimulation1994, kresseEfficiencyAbinitioTotal1996, kresseEfficientIterativeSchemes1996}. Projector-augmented wave (PAW) pseudopotentials~\cite{kresseUltrasoftPseudopotentialsProjector1999} are used, with valence configurations of H (1s$^{1}$), F (2s$^{2}$2p$^{5}$), Na (3s$^{1}$), Cl (3s$^{2}$3p$^{5}$), Rb (4s$^{2}$4p$^{6}$5s$^{1}$), and Cs (5s$^{2}$5p$^{6}$6s$^{1}$). Exchange-correlation interactions are modeled within the generalized gradient approximation using the PBEsol functional~\cite{perdewRestoringDensityGradientExpansion2008}. For bulk calculations, an $8 \times 8 \times 8$ $\Gamma$-centered $\mathbf{k}$-mesh~\cite{monkhorstSpecialPointsBrillouinzone1976} for the 8-atom conventional unit cell, together with a cutoff energy of \SI{600}{\eV}, ensures convergence of both the total energy and phonon frequencies. The bulk structures are optimized by relaxing the lattice parameters, until the total energy converges to within $10^{-6}$ \SI{}{\meV} and atomic forces to within $10^{-2}$ \SI{}{\meV\per\angstrom}.

We then use the optimized bulk structures to construct (001)-oriented slabs consisting of 11 atomic layers. A vacuum region of at least \SI{20}{\angstrom} is introduced along the $z$ direction to avoid spurious interactions between periodic images. Since the unit cell size of the slabs reduces by $\frac{1}{\sqrt{2}}$ in the in-plane directions, we use a $12 \times 12 \times 1$ $\mathbf{k}$-mesh to sample reciprocal space. The slabs are then relaxed by allowing only the atomic positions to vary, using the same convergence criteria as for the bulk structures.

Phonon calculations are performed using the finite displacement method, as implemented in \texttt{phonopy}~\cite{togoFirstprinciplesPhononCalculations2023, togoImplementationStrategiesPhonopy2023}. Interatomic force constants are determined with atomic displacements of \SI{0.01}{\angstrom}. We account for long-range dipole-dipole interactions between phonons by using the usual non-analytical correction term~\cite{gonzeInteratomicForceConstants1994, gonzeDynamicalMatricesBorn1997}. For the bulk phonons we sample a $5 \times 5 \times 5$ primitive supercell with a $3 \times 3 \times 3$ $\mathbf{k}$-mesh, whereas we use a $4 \times 4 \times 1$ supercell and a $3 \times 3 \times 1$ $\mathbf{k}$-mesh for the slab phonons. A $96 \times 96 \times 1$ $\mathbf{q}$-mesh is used for the two-dimensional cross section of the Brillouin zone. Note 1 in the Supplemental Material (SM)~\cite{supplementalMaterial} (see also Refs.~\cite{ceperleyGroundStateElectron1980, perdewGeneralizedGradientApproximation1996, straumanisGitterKonstantenNaCl1936, nickelsXRayDiffractionStudies1949, raunioPhononDispersionRelations1969, schmunkLatticeDynamicsSodium1970, breseStructuresRbDCsD1991, zintlUeberAlkalihydride1931, goldschmidtGesetzeKrystallochemie1926, uedaChiralPhononsPolar2025, grimmeEffectDampingFunction2011, braggXLIICrystallineStructure1914, westgrenXXXIXrayAnalysis1925, hasselUeberKristallstrukturGraphits1924, chisLargeSurfaceCharge2008, derouffignacLatticeDynamicsGraphite1981} therein) provides benchmarking against experiments and convergence tests.

\section*{Surface phonons}

Rocksalt-structure materials are highly symmetric, with space group $Fm\bar{3}m$ [Fig.~\ref{fig-1:bulk_phonons}(a)]. Their primitive unit cell contains only two atoms, so that their phonon dispersion comprises six phonon branches. The DFT-calculated phonon dispersion in Fig~\ref{fig-1:bulk_phonons}(b) shows doubly degenerate transverse modes and a single longitudinal mode for both the acoustic and optical phonon branches. These features persist in the calculated phonon dispersion of the 11-layer (001)-oriented NaCl slab (space group $P4/mmm$) [Fig.~\ref{fig-2:surface_phonons}(a)]. While the NaCl(001) surface is non-polar, in the sense of being uncharged, the surfaces locally break $\mathcal{P}$ and $\mathcal{P T}$ symmetry (SM Note 2~\cite{supplementalMaterial}). The broken translational symmetry along the $z$ axis results in a 22-atom slab unit cell with 66 phonon branches. These additional branches form discrete phonon subbands that emerge from the bulk phonon bands, as a consequence of confinement along the out-of-plane direction~\cite{allenStudiesVibrationalSurface1971a, kressSurfacePhonons1991}.

Beyond increasing the number of modes, the broken symmetry at the surfaces also gives rise to qualitatively new modes, which are the surface phonons, localized at or near the surface~\cite{chenSurfacePseudosurfaceModes1971, allenStudiesVibrationalSurface1971a, allenStudiesVibrationalSurface1971b, chenStudiesVibrationalSurface1977, wallisSurfaceVibrationalProperties1985, kressSurfacePhonons1991}. To quantify the degree of localization at the surface $\phi^{\mathrm{surf}}_{\mathbf{q} \nu}$ of a given mode $\nu$, we sum the projections of the atomic phonon eigenvectors $| \epsilon_{\mathbf{q} \nu, \kappa} \rangle$ at each wave vector $\mathbf{q}$ over the surface atoms $\kappa$
\begin{equation}
    \label{eq:phonon_localization}
    \phi^{\mathrm{surf}}_{\mathbf{q} \nu} = \sum_{\kappa} \langle \epsilon_{\mathbf{q} \nu, \kappa} | \epsilon_{\mathbf{q} \nu, \kappa} \rangle.
\end{equation}
Surface phonons, which reside predominantly on the surface atoms, i.e. layers 1 and 11 in Fig.~\ref{fig-2:surface_phonons}(b), have $\phi^{\mathrm{surf}}_{\mathbf{q} \nu} \approx 1$.

In Fig.~\ref{fig-2:surface_phonons}(a), we color-code our calculated phonon dispersion according to the value of $\phi_{\mathbf{q} \nu}^{\mathrm{surf}}$. We find that surface phonons, with atomic displacements strongly confined to the outermost layers, appear throughout the Brillouin zone. These modes occur at the low-frequency end of the phonon subbands due to reduced coordination of the surface atoms and they retain the character of their phonon subband~\cite{chenStudiesVibrationalSurface1977, kressSurfacePhonons1991}. For example, the most surface-localized phonon near the X point at approximately \SI{4.0}{\THz} keeps the transverse optical (TO) character of its bulk counterpart~\cite{chenStudiesVibrationalSurface1977}.

\begin{figure*}[htbp]
    \includegraphics{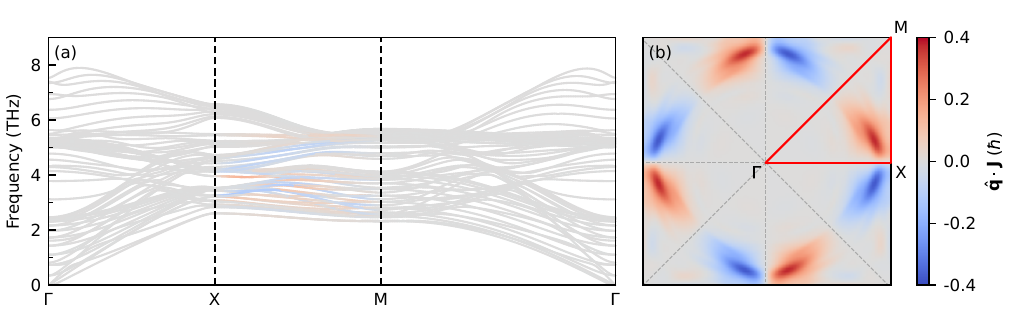}
    \caption{Chiral phonons in slab of NaCl. (a) Slab phonon dispersion, with the phonon bands colored with the phonon chirality $\hat{\mathbf{q}} \cdot \mathbf{J}_{\mathbf{q} \nu}$ of the top layer. (b) Constant-frequency cross section of the Brillouin zone at \SI{4.0}{\THz} showing the chirality of the top layer. The high-symmetry $\mathrm{\Gamma - X - M - \Gamma}$ path is shown in red. Mirror planes are indicated with dashed gray lines.}
    \label{fig-3:chiral_surface_phonons}
\end{figure*}

\section*{Chiral phonons}

Next, we analyze the chirality of these surface phonons, by first calculating their angular momentum~\cite{zhangAngularMomentumPhonons2014, zhangChiralPhononsHighSymmetry2015}, as
\begin{equation}
    \label{eq:phonon_angular_momentum}
    J^{\alpha}_{\mathbf{q} \nu} = \sum_{\kappa} J^{\alpha}_{\mathbf{q} \nu, \kappa} = \sum_{\kappa} \hbar \langle \epsilon_{\mathbf{q} \nu, \kappa} | \hat{S}^{\alpha} | \epsilon_{\mathbf{q} \nu, \kappa} \rangle.
\end{equation}
Here, $\hat{S}^{\alpha}$ is the operator for phonon circular polarization around the axis $\alpha$ ($= x, y, z$), defined as
\begin{equation}
    \label{eq:circular_polarization_operator}
    \hat{S}^{\alpha} = ( | R^{\alpha} \rangle \langle R^{\alpha} | - | L^{\alpha} \rangle \langle L^{\alpha} | ),
\end{equation}
with $| R^{\alpha} \rangle$ and $| L^{\alpha} \rangle$ the bases for right- and left-handed circular motion around axis $\alpha$, respectively. These bases are constructed from orthogonal vibrational components perpendicular to $\alpha$ with a relative phase shift of $\pm90^\circ$, corresponding to circular motion of opposite handedness. Then, as a measure of the chirality of a phonon mode, we compute the helicity of the phonon modes as $\hat{\mathbf{q}} \cdot \mathbf{J}$ with $\hat{\mathbf{q}} = \frac{\mathbf{q}}{| \mathbf{q} |}$. At $\Gamma$, $\hat{\mathbf{q}}$ is undefined because $\mathbf{q} = 0$, so we evaluate the helicity at $\Gamma$ by taking the limit $\mathbf{q} \rightarrow 0$, with $\hat{\mathbf{q}}$ defined by the corresponding limiting direction of the wave vector.

Figure~\ref{fig-3:chiral_surface_phonons}(a) shows again the 11-layer NaCl slab phonon dispersion, with the color-coding this time indicating the phonon chirality of the top surface layer. Along the high-symmetry path, only modes along the $\mathrm{X-M}$ direction are chiral, having $\hat{\mathbf{q}} \cdot \mathbf{J} \neq 0$. Interestingly, the highly surface-localized mode near the X point at \SI{4.0}{\THz} exhibits a pronounced right-handed character at the top surface ($\hat{\mathbf{q}} \cdot \mathbf{J} > 0$). If we instead consider the bottom surface, the phonon chirality reverses sign, yielding a left-handed character ($\hat{\mathbf{q}} \cdot \mathbf{J} < 0$), as shown in Fig. S3 in SM Note 2~\cite{supplementalMaterial}. This sign reversal follows directly from the inversion symmetry relating the two surfaces of the slab, such that the combined motion of both surfaces remains achiral (Fig. S3). These results demonstrate that the reduced symmetry at the crystal surface removes the symmetry constraint that forbids phonon chirality in NaCl, allowing chiral surface phonons to emerge. In SM Note 2~\cite{supplementalMaterial}, we provide a general symmetry analysis showing that, in general, the breaking of $\mathcal{P T}$ symmetry at crystal surface makes phonon chirality allowed.

To investigate the evolution of chirality across the Brillouin zone, we construct a constant-frequency cross section at 4.0 THz, chosen to correspond to the highly surface-localized mode at the X point [Fig.~\ref{fig-3:chiral_surface_phonons}(b); details are provided in SM Note 2~\cite{supplementalMaterial}]. First, we note the absence of phonon chirality along the $\mathrm{\Gamma - X}$ and $\mathrm{\Gamma - M}$ lines, both of which lie within mirror planes and are therefore invariant under reflection. Since phonon chirality $\hat{\mathbf{q}} \cdot \mathbf{J}$ changes sign under reflection, it is therefore required by symmetry to be zero along these lines~\footnote{We note that phonons along these high-symmetry lines can be cycloidal, with the ions tracing circular trajectories within a plane perpendicular to the propagation direction ($\hat{\mathbf{q}} \perp \mathbf{J}$). A full map of the phonon cycloidicity, i.e. $\hat{\mathbf{q}} \times \mathbf{J}$, throughout the Brillouin zone is provided in SM Note 2~\cite{supplementalMaterial}.}. Away from these nodal lines, we find surface phonon chirality throughout the entire Brillouin zone cross section, with the mirror symmetries of the Brillouin zone reversing the chirality of the surface phonons. 

\begin{figure}[htbp]
    \includegraphics{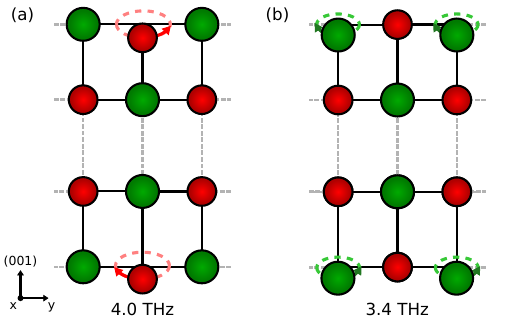}
    \caption{Atomic motion of chiral surface phonons. (a),(b) Atomic displacements at the top and bottom surfaces of the NaCl slab for chiral surface phonons along the $\mathrm{X-M}$ path at (a) \SI{4.0}{\THz} and (b) \SI{3.4}{\THz}, with wave vector $\mathbf{q} = (\frac{1}{2}, \frac{1}{10}, 0)$. The red and green spheres represent Na$^{+}$ and Cl$^{-}$ ions, respectively. Dashed gray lines indicate bonds connected through periodic boundaries or extending into the slab.}
    \label{fig-4:surface_atomic_motion}
\end{figure}

\begin{figure*}[htbp]
    \includegraphics{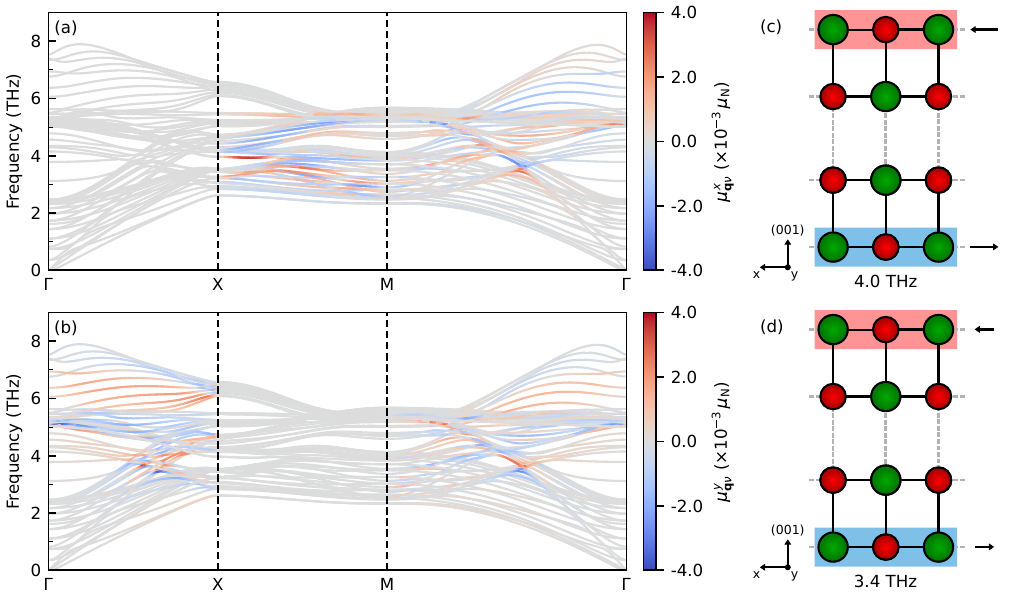}
    \caption{Phonon magnetic moments in a NaCl slab. (a),(b) Slab phonon dispersions with the phonon bands colored according to the (a) $x$ and (b) $y$ components of the phonon magnetic moment in the top surface layer. (c),(d) In-plane magnetic moments induced by chiral surface phonons at (c) \SI{4.0}{\THz} and (d) \SI{3.4}{\THz}, with wave vector $\mathbf{q} = (\frac{1}{2}, \frac{1}{10}, 0)$ along the $\mathrm{X-M}$ path. The red and green spheres represent Na$^{+}$ and Cl$^{-}$ ions, respectively. Dashed gray lines indicate bonds connected through periodic boundaries or extending into the slab and the arrows indicate the magnitude and orientation of the mode-resolved magnetic moments. Phonons with wave vector $-\mathbf{q}$ have in-plane magnetic moments oriented in the opposite direction, as required by time-reversal symmetry in the nonmagnetic slab.}
    \label{fig-5:surface_magnetization}
\end{figure*}

In Fig.~\ref{fig-4:surface_atomic_motion} we show the atomic motion in the two chiral surface modes at \SI{4.0}{\THz} and \SI{3.4}{\THz} with wave vector $\mathbf{q} = (\frac{1}{2}, \frac{1}{10}, 0)$ along the $\mathrm{X-M}$ path. For the \SI{4.0}{\THz} mode [Fig.~\ref{fig-4:surface_atomic_motion}(a)], we see that the Na$^{+}$ ions trace circular paths, while the Cl$^{-}$ ions exhibit almost no motion. In the \SI{3.4}{\THz} mode [Fig.~\ref{fig-4:surface_atomic_motion}(b)], the chirality instead stems from circulating Cl$^{-}$ ions, while the Na$^{+}$ ions remain nearly immobile. The lower frequency of this mode results from the larger mass of the Cl$^{-}$ ions compared to the Na$^{+}$ ions. For both modes, we find opposite handedness of the circular trajectories at each surface, with each surface associated with a distinct phonon chirality, while the atomic motion of the entire slab remains achiral.

\section*{Magnetic moments}

Next, we calculate the magnetic moments that arise from the chiral surface phonons. Within the picture of dynamical multiferroicity~\cite{juraschekDynamicalMultiferroicity2017, juraschekOrbitalMagneticMoments2019}, ions trace out circular (or elliptical) trajectories, effectively forming current loops that generate orbital magnetic moments, related to the phonon angular momentum by
\begin{equation}
    \label{eq:phonon_magnetic_moment}
    \mu^{\alpha}_{\mathbf{q} \nu} = \sum_{\kappa} \gamma^{\alpha \beta}_{\kappa} J^{\beta}_{\mathbf{q} \nu, \kappa}.
\end{equation}
Here $\gamma_{\kappa} = \frac{e \mathbf{Z^{*}_{\kappa}}}{2 m_{\kappa}}$ is the gyromagnetic ratio of ion $\kappa$, in terms of its Born effective charge tensor $\mathbf{Z}^{*}_{\kappa}$ and mass $m_{\kappa}$. We note that experiments on bulk systems suggest much larger phonon magnetic moments than theoretical estimates based on this expression~\cite{baydinMagneticControlSoft2022, basiniTerahertzElectricfielddrivenDynamical2024}, with the origin of the differences still under discussion. We expect a similar discrepancy for the magnetic moments associated with chiral surface phonons. Our values should therefore be regarded as a lower bound.

Figure~\ref{fig-5:surface_magnetization} presents the slab phonon dispersion, now color-coded with the $x$ [Fig.~\ref{fig-5:surface_magnetization}(a)] and $y$ components [Fig.~\ref{fig-5:surface_magnetization}(b)] of the mode-resolved phonon magnetic moments in the top layer of the slab, calculated using Eq.~(\ref{eq:phonon_magnetic_moment}). Several chiral surface phonons exhibit finite magnetic moments arising from circular ionic motion. We find that the largest magnetic moment occurs for the chiral surface phonon at \SI{4.0}{\THz} with wave vector $\mathbf{q} = (\frac{1}{2}, \frac{1}{10}, 0)$, where it originates from the circular motion of the Na$^{+}$ ions. At the same wave vector, the chiral surface phonon at \SI{3.4}{\THz} generates a smaller magnetic moment through the circular motion of the Cl$^{-}$ ions. Despite the opposite handedness of the ionic motion [Fig.~\ref{fig-4:surface_atomic_motion}], both modes result in the same orientation of the magnetic moments due to oppositely charged circulating ions. Since the ions rotate around in-plane axes, the resulting magnetic moments are likewise oriented within the plane. Moreover, the localization of chiral surface phonons at the slab surfaces, confines the magnetic moments to the surface layers [Figs.~\ref{fig-5:surface_magnetization}(c),(d)]. For both modes, the magnetic moment aligns parallel to the $x$ axis at the top surface [Figs.~\ref{fig-5:surface_magnetization}(a),(b)], while at the bottom surface it is antiparallel to it (SM Note 3~\cite{supplementalMaterial}). This opposite orientation originates from the opposite handedness of the ionic motion at the two surfaces [Fig.~\ref{fig-4:surface_atomic_motion}]. These mode-resolved magnetic moments are characteristic of the individual phonon eigenmodes and, by themselves, do not imply static surface magnetization. Since the slab is nonmagnetic, time-reversal symmetry connects the modes at $\mathbf{q}$ and $- \mathbf{q}$, enforcing opposite angular momenta and magnetic moments for the two modes. Thus, a net surface magnetic response can arise only when the populations of the modes with opposite magnetic moments are brought out of equilibrium, for example through selective excitation or field gradients.

Finally, we note that in addition to the $\mathrm{X - M}$ path in reciprocal space, we also find surface phonons with finite magnetic moments along the $\mathrm{\Gamma - X}$ and $\mathrm{\Gamma - M}$ paths [Figs.~\ref{fig-5:surface_magnetization}(a),(b)]. While these modes are not chiral, they are cycloidal and the atoms trace out circular paths, producing finite magnetic moments. As was the case for the chiral surface phonons, the opposite handedness of the atomic motion at the top and bottom surface leads to oppositely oriented magnetic moments.

\section*{Other materials}

To illustrate the generality of our findings, we calculate the vibrational properties of the bulk and slabs of other rocksalt-structure compounds, as well as other types of materials. The rocksalt-structure compounds RbF and CsH exhibit qualitatively similar behavior to NaCl, due to their identical crystal structures. In both compounds, the larger mass difference between the A$^{+}$ cation and B$^{-}$ anion leads to a more pronounced acoustic-optical phonon gap, when compared to NaCl. Nevertheless, in all three compounds, the presence of surfaces gives rise to surface-localized phonon modes, some of which exhibit chiral ionic motion, resulting in a finite phonon angular momentum and associated surface magnetism.

Beyond rocksalt-structure compounds, we also investigate the surface phonons of copper and graphite. Despite their distinct crystal structures and bonding characteristics, both materials exhibit chiral surface-localized phonons. Note, however, that since both materials are monoatomic, their chiral surface phonons have zero magnetic moment. These results demonstrate that chiral surface phonons are not unique to rocksalt-structure materials and support our symmetry analysis, while highlighting the role of material-specific chemistry in determining their magnitude. Details of the (chiral) surface phonons and their properties can be found in SM Note 4~\cite{supplementalMaterial} for RbF and CsH. The corresponding results for Cu(111) and (0001)-oriented graphite slabs are presented in SM Note 5~\cite{supplementalMaterial}.

\section*{Discussion}

Our results show that chiral phonons are universal to crystal surfaces, with their character and magnitude determined by the microscopic details of the surface. These chiral surface phonons carry finite phonon angular momentum and often have surface magnetic moments, suggesting that surface measurements sensitive to such effects should take this into account.

In our slab calculations, the two surfaces are related by inversion symmetry, and therefore exhibit opposite phonon angular momentum, chirality and magnetic moments. Consequently, their contributions cancel in the net response of the slab. This cancellation can be avoided by either resolving the two surfaces individually or by considering structures with inequivalent surfaces, such as films supported on substrates or slabs with different surface terminations.

In experiments, the predicted chiral surface phonons could potentially be probed using resonant inelastic X-ray scattering (RIXS), which has recently been used to measure chiral phonons in bulk materials~\cite{uedaChiralPhononsQuartz2023, uedaChiralPhononsPolar2025, grimesElectricfieldSwitchingGwave2026}. A grazing-incidence geometry may enhance the sensitivity to surface-localized modes, while the incident X-ray polarization could provide access to their chiral character.

An interesting connection can also be made to the phonon angular momentum Hall effect~\cite{parkPhononAngularMomentum2020, lopezAtomisticTheoryPhonon2026}, although the physical origins of the two phenomena are distinct. The phonon angular momentum Hall effect is a non-equilibrium transport phenomenon in which a temperature gradient drives an accumulation of phonon angular momentum at the surface, whereas the modes considered here are surface-localized eigenmodes that intrinsically carry angular momentum. Their coherent excitation therefore provides an alternative route to generating surface phonon angular momentum and associated magnetic moments with well-defined polarization, without relying on a thermal transport process.

\section*{Acknowledgements}

We thank Dominik Juraschek, Urs Staub, and Martina Basini for useful discussions. This work was supported by the Swiss National Science Foundation (SNF) under Grant No. 225790 and by ETH Z\"{u}rich. Calculations were performed on the Swiss National Supercomputing Centre (CSCS) Daint cluster under Project No. lp61 and on the ETH Z\"{u}rich Euler cluster.

\section*{Data availability}

The data that support the findings of this article are openly available \cite{zenodoData2026}.

\bibliography{references}

\end{document}


\title{Supplemental Material: \\ Chiral Surface Phonons}

\author{Mike Pols}
\email{mike.pols@mat.ethz.ch}
\affiliation{%
    Materials Theory, ETH Zurich, CH-8093 Z\"urich, Switzerland
}%

\author{Nicola A. Spaldin}
\email{nicola.spaldin@mat.ethz.ch}
\affiliation{%
    Materials Theory, ETH Zurich, CH-8093 Z\"urich, Switzerland
}%

\date{\today}

\maketitle

\clearpage

\tableofcontents

\clearpage

\section{Computational methods}

\subsection*{Exchange-correlation (XC) functionals}

To determine the most suitable exchange-correlation (XC) functional for our phonon calculations, we evaluate how well they reproduce both experimental geometries and phonon frequencies. Our analysis includes the LDA~\cite{ceperleyGroundStateElectron1980}, PBE~\cite{perdewGeneralizedGradientApproximation1996}, and PBEsol~\cite{perdewRestoringDensityGradientExpansion2008} functionals. For each functional, we optimize the geometry of NaCl, CsH, and RbF, followed by a determination of the transverse optical (TO) mode frequency at the $\Gamma$ point. The resulting lattice parameters $a$ and TO mode frequencies $\nu_{\mathrm{TO}}$ are shown in Table~\ref{tab:computational_rocksalt}.

\begin{table*}[htbp!]
\caption{\label{tab:computational_rocksalt}Lattice parameters $a$ and transverse optical (TO) mode frequencies $\nu_{\mathrm{TO}}$ of rocksalt compounds with $Fm\bar{3}m$ space group (\#225) from density functional theory (DFT) calculations and experiments.}
\begin{ruledtabular}
    \begin{tabular}{cccc}
    Compound & XC functional & $a$ (\SI{}{\angstrom}) & $\nu_{\mathrm{TO}}$ (\SI{}{\THz}) \\
    \colrule
    NaCl    & LDA           & 5.43      &  5.79     \\
            & PBE           & 5.65      &  4.61     \\
            & PBEsol        & 5.56      &  5.12     \\
            & Experiment    & 5.63~\cite{straumanisGitterKonstantenNaCl1936, nickelsXRayDiffractionStudies1949} & 5.0~\cite{raunioPhononDispersionRelations1969, schmunkLatticeDynamicsSodium1970} \\ \colrule
    CsH     & LDA           & 6.11      & 13.54     \\
            & PBE           & 6.44      & 10.56     \\
            & PBEsol        & 6.26      & 12.03     \\
            & Experiment    & 6.37~\cite{breseStructuresRbDCsD1991}, 6.38~\cite{zintlUeberAlkalihydride1931} & - \\ \colrule
    RbF     & LDA           & 5.47      &  5.21     \\
            & PBE           & 5.74      &  3.99     \\
            & PBEsol        & 5.61      &  4.54    \\
            & Experiment    & 5.64~\cite{goldschmidtGesetzeKrystallochemie1926} & - \\
    \end{tabular}
\end{ruledtabular}
\end{table*}

Focusing on the lattice parameters first, we find that LDA underestimates the lattice parameters, while both PBE and PBEsol predict lattice parameters much closer to experiments. When examining the mode frequencies, however, only PBEsol accurately captures the experimentally observed TO mode frequency of NaCl. Therefore, we use PBEsol as the XC functional for our phonon calculations in rocksalt compounds, providing the best agreement with experimental geometries and frequencies.

\subsection*{Convergence of slab thickness}

\begin{figure}[htbp]
    \includegraphics{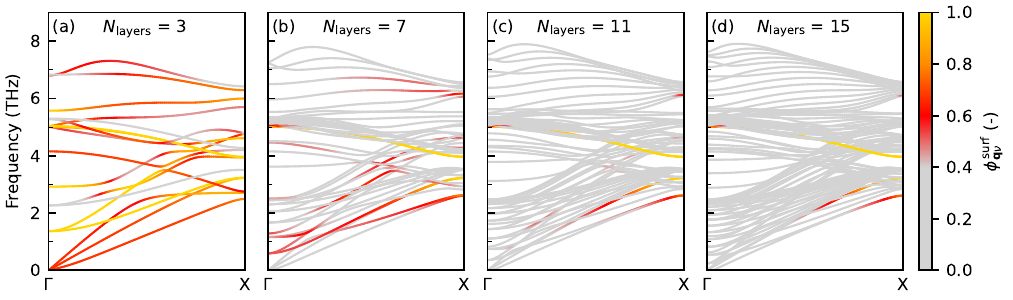}
    \caption{Convergence of phonon dispersion with slab thickness $N_{\mathrm{layers}}$. Phonon dispersions of (a) 3-layer, (b) 7-layer, (c) 11-layer, and (d) 15-layer NaCl slab. The phonon bands are colored with the surface localization $\phi_{\mathbf{q} \nu}^{\mathrm{surf}}$.}
    \label{si-fig:slab_thickness}
\end{figure}

To investigate the required slab thickness for convergence of the surface phonons, we calculate the phonon dispersion for a series of NaCl slabs with different thicknesses. The resulting dispersions, for $N_{\mathrm{layers}}$ = 3, 7, 11, and 15 layers, are shown in Fig.~\ref{si-fig:slab_thickness}. We see that surface states become progressively more localized in reciprocal space as the slab thickness increases. Notably, only minimal changes in the character of the phonon bands are observed when increasing the number of layers in the rocksalt slabs from $N_{\mathrm{layers}} = 11$ to $N_{\mathrm{layers}} = 15$. As such, we conclude that slabs consisting of 11 layers are sufficient for investigating the surface phonons of rocksalt compounds.

\subsection*{Convergence of vacuum thickness}

\begin{figure}[htbp]
    \includegraphics{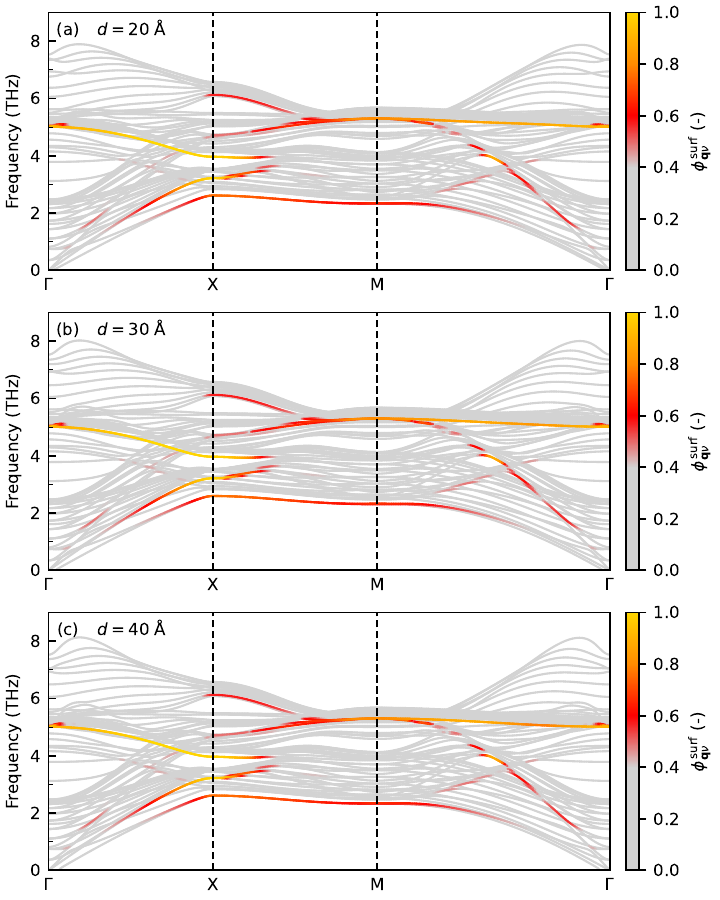}
    \caption{Convergence of phonon dispersion with vacuum thickness. Phonon dispersions of an 11-layer NaCl slab with (a) \SI{20}{\angstrom}, (b) \SI{30}{\angstrom}, and (c) \SI{40}{\angstrom} vacuum. The phonon bands are colored with the surface localization $\phi_{\mathbf{q} \nu}^{\mathrm{surf}}$.}
    \label{si-fig:vacuum_thickness}
\end{figure}

Next, we assess the convergence of the surface phonon calculations with respect to the vacuum thickness by computing the phonon dispersions of 11-layer NaCl slabs separated by vacuums of \SI{20}{\angstrom}, \SI{30}{\angstrom}, and \SI{40}{\angstrom}. The resulting phonon dispersions are shown in Fig.~\ref{si-fig:vacuum_thickness}. We find that the surface phonons are essentially unchanged as the vacuum thickness is increased, exhibiting only minor frequency shifts of approximately \SI{0.02}{\THz}. This demonstrates that a vacuum thickness of \SI{20}{\angstrom} yields sufficiently converged phonon dispersions for our analysis.

\clearpage

\section{Chiral phonons}

\subsection*{Symmetry analysis}

We now analyze the symmetry properties of a phonon characterized by wave vector $\mathbf{q}$ and a phonon angular momentum $\mathbf{J}^{\mathrm{ph}}$, closely following the symmetry analysis for bulk systems in Refs.~\cite{cohClassificationMaterialsPhonon2023, yangSymmetryguidedCatalogueChiral2026}. Under spatial inversion ($\mathcal{P}$), the wave vector changes sign, whereas the phonon angular momentum remains unchanged:
\begin{equation}
    \label{eq:p_operation}
    ( \mathbf{q}, \mathbf{J}^{\mathrm{ph}} ) \xrightarrow[]{\mathcal{P}} ( - \mathbf{q}, \mathbf{J}^{\mathrm{ph}} ).
\end{equation}
In contrast, under time-reversal symmetry ($\mathcal{T}$), both the wave vector and phonon angular momentum change sign:
\begin{equation}
    \label{eq:t_operation}
    ( \mathbf{q}, \mathbf{J}^{\mathrm{ph}} ) \xrightarrow[]{\mathcal{T}} ( - \mathbf{q}, - \mathbf{J}^{\mathrm{ph}} ).
\end{equation}
When both operations are combined into the $\mathcal{P T}$ operation, the wave vector remains unchanged while the direction of the phonon angular momentum is reversed:
\begin{equation}
    \label{eq:pt_operation}
    ( \mathbf{q}, \mathbf{J}^{\mathrm{ph}} ) \xrightarrow[]{\mathcal{P T}} ( \mathbf{q}, - \mathbf{J}^{\mathrm{ph}} ).
\end{equation}
This implies that, in a system that preserves $\mathcal{P T}$ symmetry, phonons can not have a finite angular momentum, i.e. $\mathbf{J}^{\mathrm{ph}} ( \mathbf{q} ) = 0$. In particular, the phonon chirality, defined as the projection of the phonon angular momentum onto the direction of propagation, $\hat{\mathbf{q}} \cdot \mathbf{J}^{\mathrm{ph}} ( \mathbf{q} )$, is odd under $\mathcal{P T}$ operation, and is required to vanish in a system with $\mathcal{P T}$ symmetry.

For bulk crystals, this means that the bulk space group must break $\mathcal{P T}$ symmetry for chiral phonons to be allowed. For surface phonons, however, the relevant symmetry is that of the crystal surface. The force constant matrix that describes the atomic motion of the atoms at the surface is invariant under the symmetry operations of the surface space group.

Next, we will show that all surface space groups break $\mathcal{P T}$ symmetry and thus that all surfaces can host chiral phonons. Consider a crystal with a bulk space group $\mathbf{G} = \left \{ \left( \mathbf{R} | \mathbf{t} \right) \right \}$, where $\left \{ \mathbf{R} \right \}$ is the set of point group operations and $\left \{ \mathbf{t} \right \}$ the set of fractional Bravais lattice translations. Upon creating a surface along the $( hkl )$ Miller plane, with a surface normal vector $\hat{\mathbf{n}}$, the relevant symmetry is reduced to the surface space group~\cite{weberSurfaceMagnetizationAntiferromagnets2024}
\begin{equation}
    \label{eq:surface_space_group}
    \mathbf{G}^{\hat{\mathbf{n}}} = \left \{ \left( \mathbf{R}^{\hat{\mathbf{n}}} | \mathbf{t}^{\perp \hat{\mathbf{n}}} \right) \right \}.
\end{equation}
The elements of $\mathbf{G}^{\hat{\mathbf{n}}}$ satisfy the following condition
\begin{equation}
    \label{eq:surface_space_group_constraints}
    \left( \mathbf{R}_{i} | \mathbf{t}_{i} \right) \in \mathbf{G}^{\hat{\mathbf{n}}} \Leftrightarrow \mathbf{R}_{i} \hat{\mathbf{n}} = \hat{\mathbf{n}} \text{ and } \mathbf{t}_{i} \perp \hat{\mathbf{n}},
\end{equation}
that is, the surface space group consists of all symmetry operations that leave the surface normal invariant (up to translations parallel to the surface).

Since the inversion operation reverses the direction of the surface normal, $\mathcal{P} \hat{\mathbf{n}} = - \hat{\mathbf{n}}$, inversion cannot be a symmetry operation of an individual surface. Consequently, even when the bulk crystal is centrosymmetric, any individual crystal surface is non-centrosymmetric. The same argument applies to $\mathcal{P T}$, since the time-reversal operation does not change the spatial coordinates. Under the $\mathcal{P T}$ operation, the crystal on one side of the surface is therefore mapped onto the opposite side, which is vacuum for an isolated surface. Thus, $\mathcal{P T}$ symmetry cannot be a symmetry of an isolated crystal surface, irrespective of whether the bulk crystal is magnetic or nonmagnetic. This contrasts with bulk crystals, where magnetic order can, in specific cases such as Cr$_{2}$O$_{3}$~\cite{cohClassificationMaterialsPhonon2023}, preserve $\mathcal{P T}$ symmetry even when $\mathcal{P}$ and $\mathcal{T}$ are individually broken.

The absence of $\mathcal{P T}$ symmetry at an isolated surface removes the symmetry constraint that forbids finite phonon angular momentum and phonon chirality in the bulk. Surface phonons can therefore have finite chirality and carry finite phonon angular momentum. Whether the phonon chirality is non-zero for a given wave vector and what pattern it has throughout the Brillouin zone is determined by the remaining symmetries of the surface.

\clearpage

\subsection*{Brillouin zone cross section}

To calculate the constant-frequency cross section of the Brillouin zone, we average the phonon chirality over the phonon bands by weighting the contribution of each band with a Gaussian distribution centered around a target frequency $f_{0}$. This approach, which follows the method described in Ref.~\cite{uedaChiralPhononsPolar2025}, enables the computation of an average phonon angular momentum and, consequently, an average phonon chirality for each $\mathbf{q}$-point. The average phonon angular momentum $\langle \mathbf{J} \rangle_{\mathbf{q}}$ is computed as
\begin{equation}
    \label{eq:average_angular_momentum}
    \langle \mathbf{J} \rangle_{\mathbf{q}} = \sum_{\nu} \mathbf{J}_{\mathbf{q} \nu} \exp \left[ - \left( f_{\mathbf{q} \nu} - f_{0} \right)^{2} / \left( 2 \sigma_{f}^{2} \right) \right],
\end{equation}
where $f_{\mathbf{q} \nu}$ is the frequency of a phonon mode and $\sigma_{f}$ is the standard deviation of the Gaussian distribution. For the Gaussian weighting, we use a target frequency of $f_{0} = \SI{4.0}{\THz}$ with a standard deviation of $\sigma_{f} = \SI{0.05}{\THz}$. Then, the average phonon chirality is computed as $\hat{\mathbf{q}} \cdot \langle \mathbf{J} \rangle_{\mathbf{q}}$.

\subsection*{Handedness of surfaces}

\begin{figure*}[htbp]
    \includegraphics{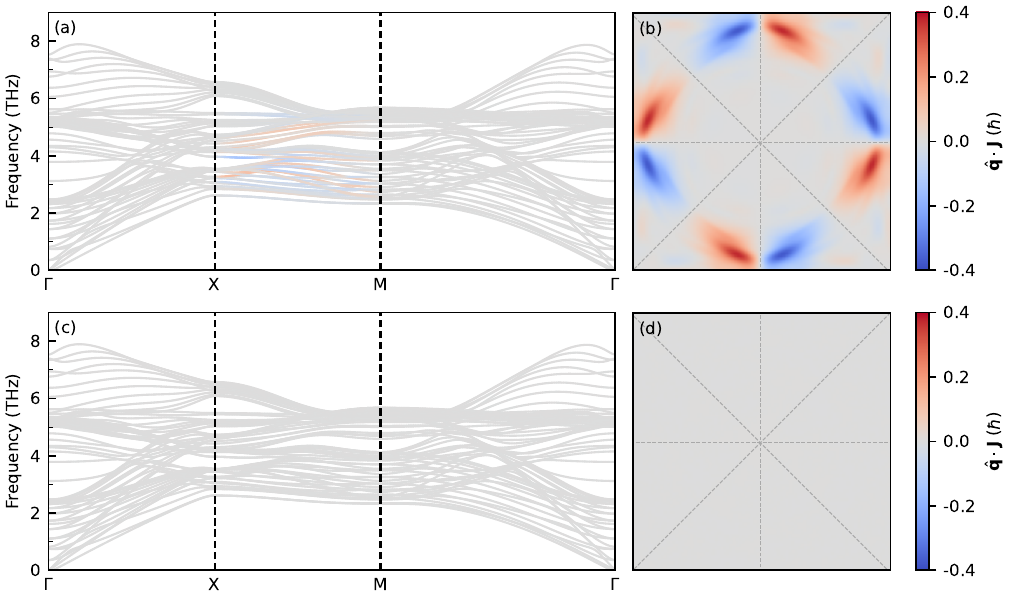}
    \caption{(a) Slab phonon dispersion, with the phonon bands colored with the phonon chirality $\hat{\mathbf{q}} \cdot \mathbf{J}_{\mathbf{q} \nu}$ of the bottom layer. (b) Constant-frequency cross section of the Brillouin zone at \SI{4.0}{\THz} showing the chirality of the top layer. The phonon chirality in the bottom layer is opposite to that of the top layer as shown in the main text in Fig. 3. (c) Slab phonon dispersion, with the phonon bands colored with the phonon chirality $\hat{\mathbf{q}} \cdot \mathbf{J}_{\mathbf{q} \nu}$ of both surface layers combined. (d) Constant-frequency cross section of the Brillouin zone at \SI{4.0}{\THz} showing the chirality of both surface layers combined. Mirror planes are indicated with dashed gray lines.}
    \label{si-fig:chiral_surface_phonons}
\end{figure*}

To illustrate that the phonon modes of each of the surfaces have opposite handedness, we show the phonon chirality of the bottom surface layer and both surface layers combined in Fig.~\ref{si-fig:chiral_surface_phonons}. The phonon chirality of both the phonon dispersion [Fig.~\ref{si-fig:chiral_surface_phonons}(a)] and Brillouin zone cross section [Fig.~\ref{si-fig:chiral_surface_phonons}(b)] of the bottom layer are opposite to those of the top layer of the slab shown in Fig. 3. The opposite handedness of the two surfaces is further confirmed by the sum of the phonon chirality across both surfaces [Fig.~\ref{si-fig:chiral_surface_phonons}(c),(d)], which sums to $\hat{\mathbf{q}} \cdot \mathbf{J} = 0$, indicating that the entire slab is achiral.

\FloatBarrier

\subsection*{Localization of chirality}

\begin{figure*}[htbp]
    \includegraphics{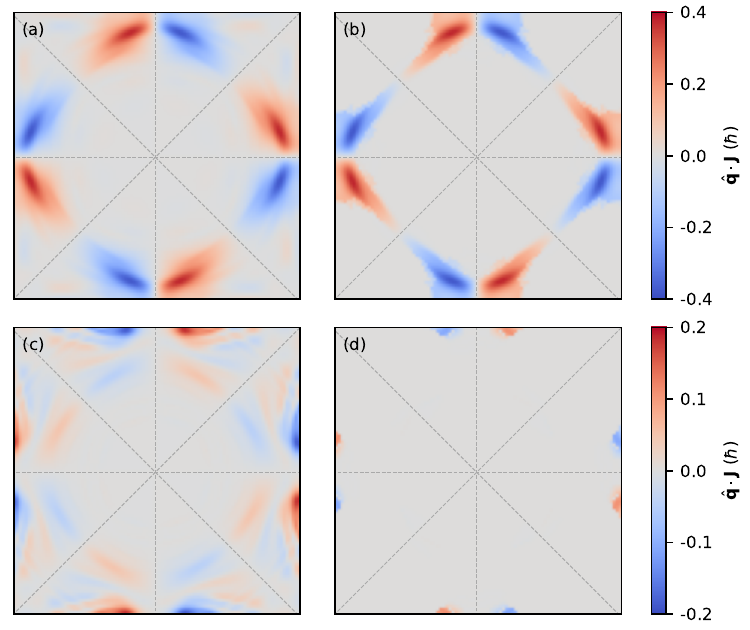}
    \caption{Surface localization of phonon chirality. Constant-frequency cross section at \SI{4.0}{\THz} for (a) all phonon modes and (b) only phonon modes with a surface localization of $\phi^{\mathrm{surf}}_{\mathbf{q} \nu} > 0.5$. Constant-frequency cross section at \SI{3.4}{\THz} for (c) all phonon modes and (d) only phonon modes with a surface localization of $\phi^{\mathrm{surf}}_{\mathbf{q} \nu} > 0.5$.}
    \label{si-fig:chiral_localization}
\end{figure*}

To confirm that the chirality that we observe in the constant-frequency cross sections stems from surface-localized phonon modes, we show the cross sections at \SI{4.0}{\THz} and \SI{3.4}{\THz} resulting from all phonon modes, compared to those with a large surface localization ($\phi^{\mathrm{surf}}_{\mathbf{q} \nu} > 0.5$) in Fig.~\ref{si-fig:chiral_localization}. For both the cross section at \SI{4.0}{\THz} [Fig.~\ref{si-fig:chiral_localization}(a),(b)] and \SI{3.4}{\THz} [Fig.~\ref{si-fig:chiral_localization}(c),(d)] we see that the overall shape remains the same close to the $\mathrm{X - M}$ path. Therefore, we can conclude that the surface phonons are the dominant contributions to the chirality we observe.

\subsection*{Convergence of vacuum thickness}

\begin{figure}[htbp]
    \includegraphics{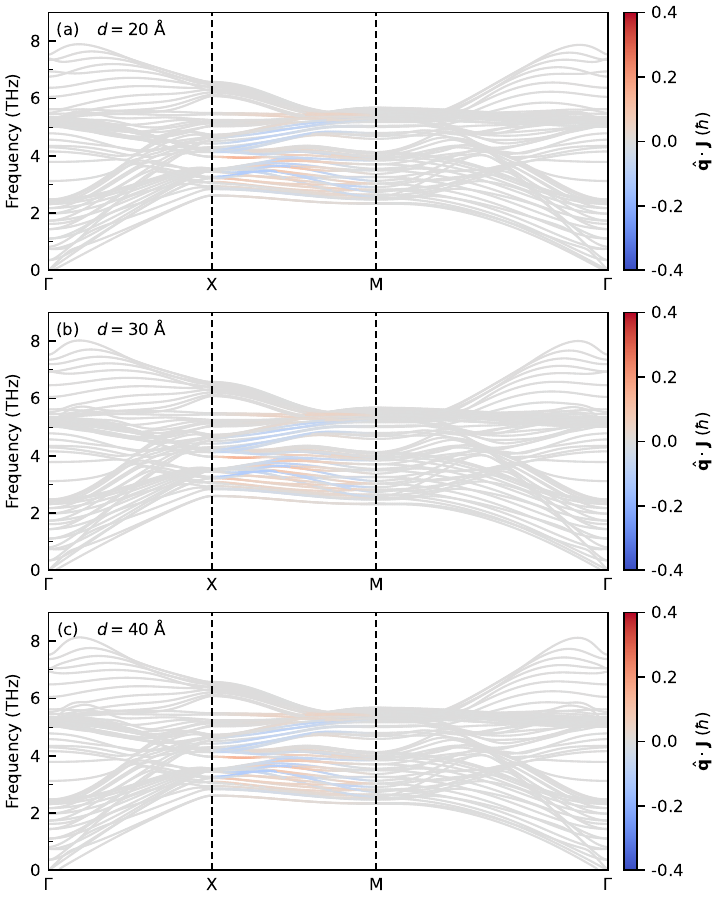}
    \caption{Convergence of chiral phonon dispersion with vacuum thickness. Phonon dispersions of an 11-layer NaCl slab with (a) \SI{20}{\angstrom}, (b) \SI{30}{\angstrom}, and (c) \SI{40}{\angstrom} vacuum. The phonon bands are colored with the phonon chirality $\hat{\mathbf{q}} \cdot \mathbf{J}_{\mathbf{q} \nu}$.}
    \label{si-fig:vacuum_thickness_chiral}
\end{figure}

Having established the convergence of the surface phonons with respect to the vacuum thickness of an 11-layer NaCl slab, we now examine whether the phonon chirality is also converged. We use the same slab calculations with vacuum thicknesses \SI{20}{\angstrom}, \SI{30}{\angstrom}, and \SI{40}{\angstrom}. The resulting phonon dispersions, colored according to their phonon chirality, are shown in Fig.~\ref{si-fig:vacuum_thickness_chiral}. Similar to the surface localization, the phonon chirality is already converged for a vacuum thickness of \SI{20}{\angstrom}, with no appreciable changes upon further increasing the vacuum thickness. This demonstrates that a \SI{20}{\angstrom} vacuum is also sufficient to converge the phonon chirality.

\clearpage

\subsection*{Cycloidicity of phonons}

\begin{figure*}[htbp]
    \includegraphics{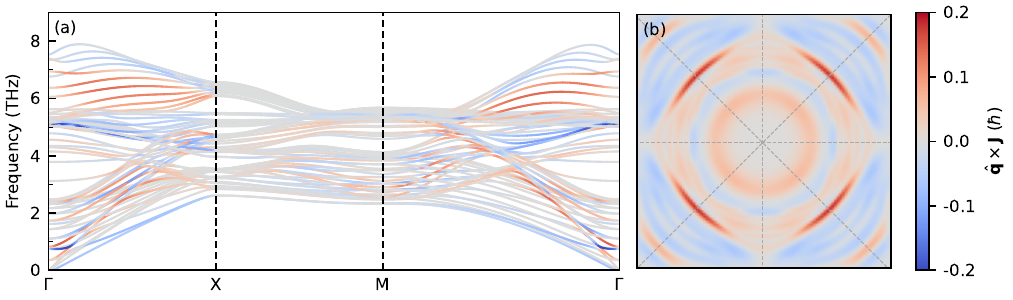}
    \caption{(a) Slab phonon dispersion, with the phonon bands colored with the phonon cycloidicity $\hat{\mathbf{q}} \times \mathbf{J}_{\mathbf{q} \nu}$ of the top layer. (b) Constant-frequency cross section of the Brillouin zone at \SI{4.0}{\THz} showing the cycloidicity of the top layer. Mirror planes are indicated with dashed gray line.}
    \label{si-fig:cycloidal_surface_phonons}
\end{figure*}

Cycloidal phonons have a finite angular momentum $\mathbf{J}_{\mathbf{q} \nu} \neq 0$ that is perpendicular to the wave vector, i.e. $\hat{\mathbf{q}} \perp \mathbf{J}$. To demonstrate that the surface layers of the rocksalt slabs have cycloidal phonon modes, we show a phonon dispersion, of which the phonon bands are color-coded with the cycloidicity of the top layer in these phonon modes $\hat{\mathbf{q}} \times \mathbf{J}$, in Fig.~\ref{si-fig:cycloidal_surface_phonons}. As shown by the phonon dispersion [Fig.~\ref{si-fig:cycloidal_surface_phonons}(a)] and Brillouin zone cross section [Fig.~\ref{si-fig:cycloidal_surface_phonons}(b)], the phonons along all high-symmetry paths exhibit some cycloidicity; even the modes along the $\mathrm{X - M}$ path have some angular momentum component perpendicular to the phonon wave vector.

\begin{figure*}[htbp]
    \includegraphics{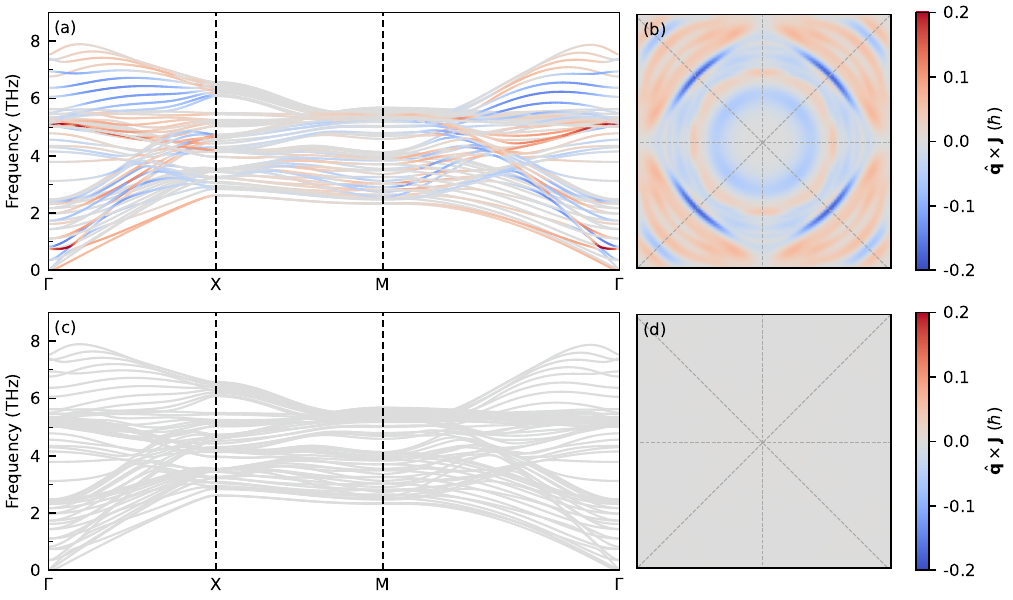}
    \caption{(a) Slab phonon dispersion, with the phonon bands colored with the phonon cycloidicity $\hat{\mathbf{q}} \times \mathbf{J}_{\mathbf{q} \nu}$ of the bottom layer. (b) Constant-frequency cross section of the Brillouin zone at \SI{4.0}{\THz} showing the cycloidicity of the bottom layer. (c) Slab phonon dispersion, with the phonon bands colored with the phonon cycloidicity $\hat{\mathbf{q}} \times \mathbf{J}_{\mathbf{q} \nu}$ of both surface layers combined. (d) Constant-frequency cross section of the Brillouin zone at \SI{4.0}{\THz} showing the cycloidicity of both surface layers combined. Mirror planes are indicated with dashed gray line.}
    \label{si-fig:cycloidal_surface_phonons_combined}
\end{figure*}

Next, we show the cycloidicity of the bottom surface and the sum over both surfaces in Fig.~\ref{si-fig:cycloidal_surface_phonons_combined}, to investigate the relationship between the cycloidicity of the surfaces. This illustrates that the cycloidicity of the bottom surface [Fig.~\ref{si-fig:cycloidal_surface_phonons_combined}(a),(b)] is exactly opposite of that in the top surface [Fig.~\ref{si-fig:cycloidal_surface_phonons}]. As was the case for the chirality, the sum of the cycloidicity over both the top and bottom surface results in a net zero cycloidicity [Fig.~\ref{si-fig:cycloidal_surface_phonons_combined}(c),(d)].

\clearpage

\section{Magnetic moments}

\subsection*{Magnetic moments of bottom surface}

\begin{figure*}[htbp]
    \includegraphics{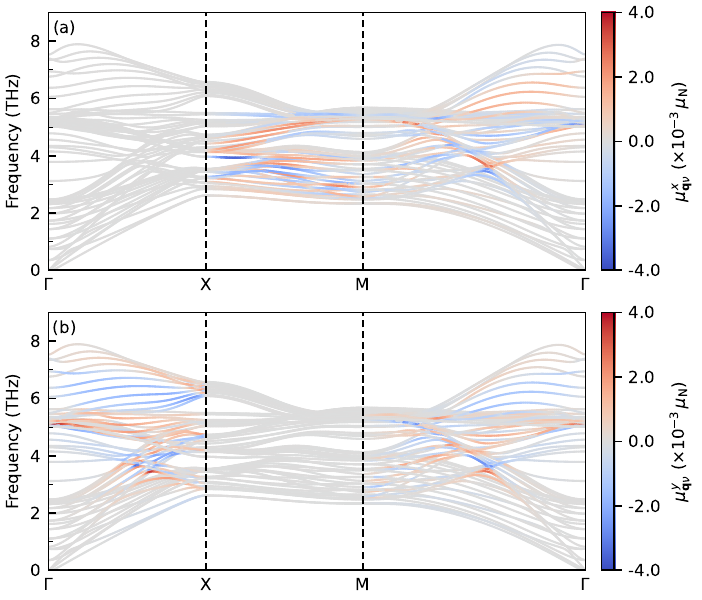}
    \caption{Slab phonon dispersion, with the phonon bands colored with the (a) $x$ and (b) $y$ components of the phonon magnetic moment resulting from the bottom surface layer.}
    \label{si-fig:magnetism_bottom_surface}
\end{figure*}

In Fig.~\ref{si-fig:magnetism_bottom_surface} the phonon magnetic moments of the bottom surface of the rocksalt slab are shown. Compared to the top layer of the slab, shown in Fig. 5, the bottom layer of the slab exhibits a phonon magnetic moment antiparallel to it: a positive $x$ component in the top surface layers corresponds to a negative $x$ component in the bottom surface layer.

\subsection*{Total magnetic moments of surfaces}

\begin{figure*}[htbp]
    \includegraphics{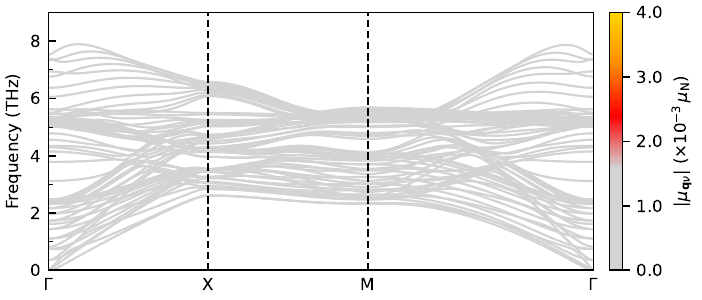}
    \caption{Slab phonon dispersion, with the phonon bands colored with the magnitude of the sum of the phonon magnetic moment over the top and bottom surface layers.}
    \label{si-fig:magnetism_total}
\end{figure*}

As a result of the opposite orientation of the magnetic moments at both surfaces, the sum over these yields a phonon magnetic moment of zero. In Fig.~\ref{si-fig:magnetism_total}, it is demonstrated that the sum of the magnetic moments over both surfaces is indeed zero.

\clearpage

\section{Other rocksalt-structure materials}

\subsection*{Bulk phonons}

\begin{figure}[htbp]
    \includegraphics{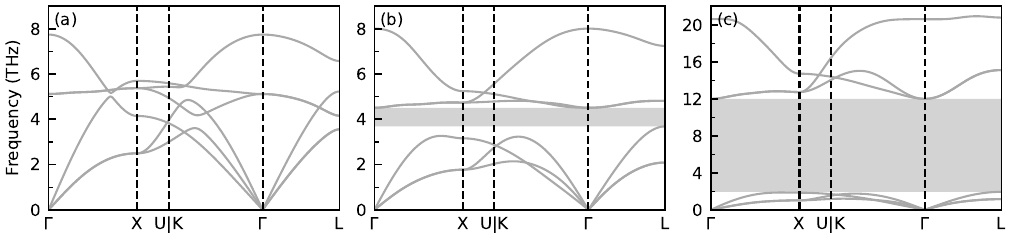}
    \caption{Phonons in AB rocksalt compounds. Bulk phonons in (a) NaCl, (b) RbF, and (c) CsH. The acoustic-optical phonon gap in RbF and CsH is indicated in gray.}
    \label{si-fig:rocksalt_bulk_phonons}
\end{figure}

Figure~\ref{si-fig:rocksalt_bulk_phonons} shows the phonon dispersion of three different compounds with a rocksalt structure: NaCl, RbF, and CsH. We see that the dispersion of the three materials are similar, with the most notable difference stemming from the mass difference between the A$^{+}$ cations and B$^{-}$ anions; the larger the mass ratio ($r = \frac{m_{\mathrm{A^{+}}}}{m_{\mathrm{B^{-}}}}$) between the two ions, the bigger the acoustic-optical phonon gap (shaded in gray).  It is absent for NaCl ($r_{\mathrm{NaCl}} = 0.65$) [Fig.~\ref{si-fig:rocksalt_bulk_phonons}(a)], and progressively larger for RbF ($r_{\mathrm{RbF}} = 4.50$) [Fig.~\ref{si-fig:rocksalt_bulk_phonons}(b)] and CsH ($r_{\mathrm{CsH}} = 131.86$) [Fig.~\ref{si-fig:rocksalt_bulk_phonons}(c)].

\subsection*{Surface phonons}

\begin{figure}[htbp]
    \includegraphics{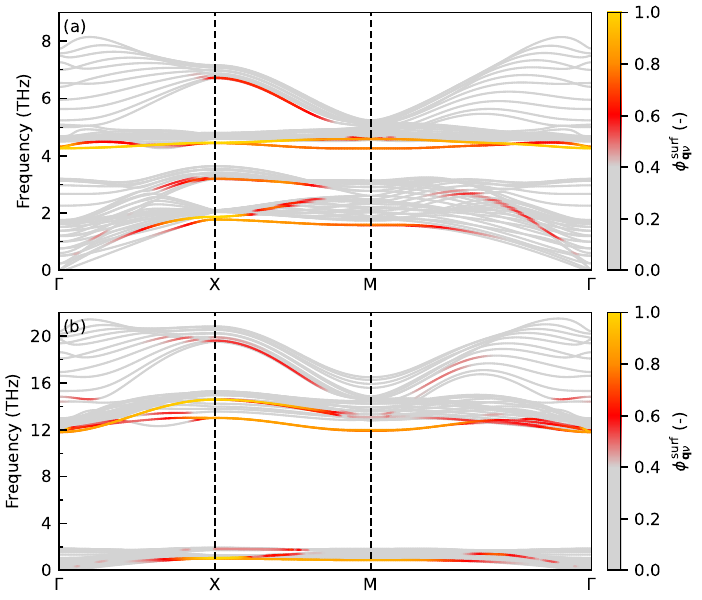}
    \caption{Surface phonons in rocksalt compounds. Slab phonon dispersion in (a) RbF and (b) CsH. The phonon bands are colored with the surface localization $\phi^{\mathrm{surf}}_{\mathbf{q} \nu}$.}
    \label{si-fig:rocksalt_surface_phonons}
\end{figure}

Next, we demonstrate that the emergence of surface phonons is not unique to NaCl, by calculating the vibrational properties of RbF and CsH slabs [Fig.~\ref{si-fig:rocksalt_surface_phonons}]. For both materials, we observe highly surface-localized phonon modes, similar to those in NaCl, albeit at shifted energies, due to different ionic masses.

\FloatBarrier

\subsection*{Chiral phonons}

\begin{figure}[htbp]
    \includegraphics{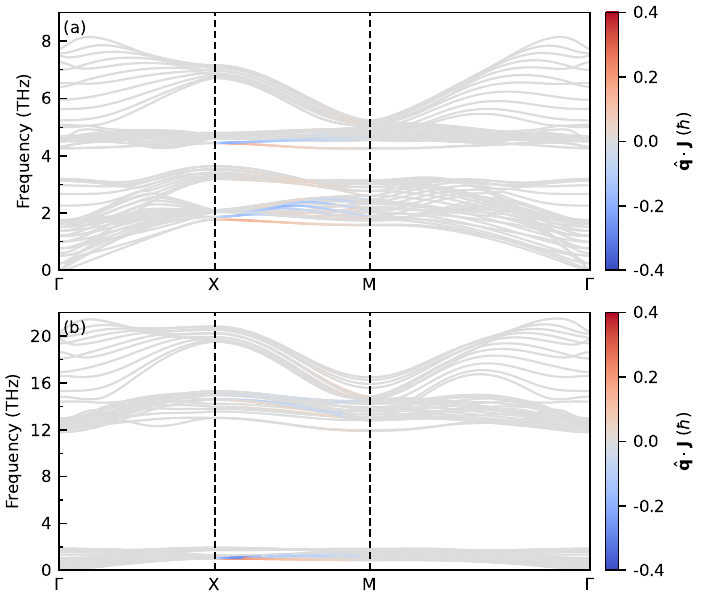}
    \caption{Chiral phonons in rocksalt compounds. Slab phonon dispersion in (a) RbF and (b) CsH. The phonon bands are colored with the phonon chirality $\hat{\mathbf{q}} \cdot \mathbf{J}_{\mathbf{q} \nu}$.}
    \label{si-fig:rocksalt_chiral_phonons}
\end{figure}

In Fig.~\ref{si-fig:rocksalt_chiral_phonons} we show the phonon dispersion of the rocksalt slabs of RbF and CsH, with the color-coding indicating the phonon chirality. It shows that the phonon modes, and in particular the surface phonons along the $\mathrm{X - M}$ path, in both compounds are chiral. As a result of the shifted surface-localized modes, the chiral surface phonons are also found at different frequencies in both materials.

\FloatBarrier

\subsection*{Magnetic moments}

\begin{figure}[htbp]
    \includegraphics{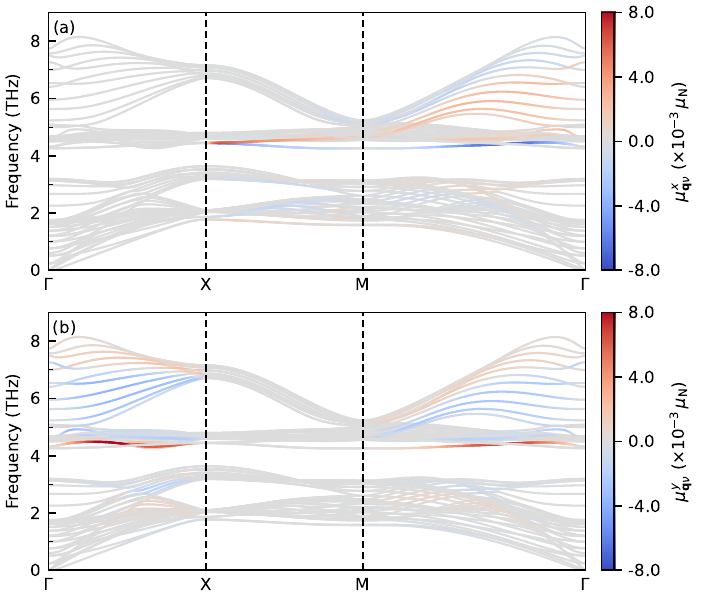}
    \caption{Phonon magnetic moments in RbF. Slab phonon dispersion, with the phonon bands colored with the (a) $x$ and (b) $y$ components of the phonon magnetic moments.}
    \label{si-fig:rocksalt_magnetism_phonons_rbf}
\end{figure}

\begin{figure}[htbp]
    \includegraphics{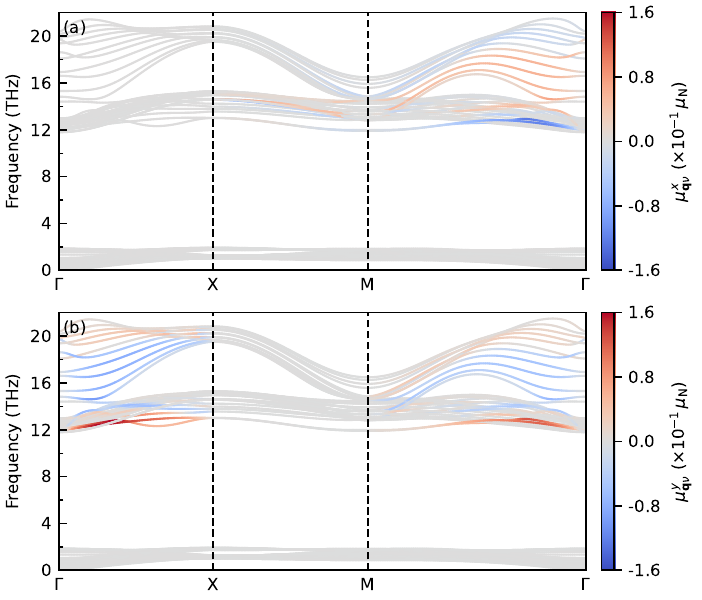}
    \caption{Phonon magnetic moments in CsH. Slab phonon dispersion, with the phonon bands colored with the (a) $x$ and (b) $y$ components of the phonon magnetic moments.}
    \label{si-fig:rocksalt_magnetism_phonons_csh}
\end{figure}

Finally, we show that the phonon modes of both compounds yield phonon magnetic moments. The phonon dispersions of RbF and CsH, color-coded with the phonon magnetic moments of the top surface layer, are shown in Fig.~\ref{si-fig:rocksalt_magnetism_phonons_rbf} and Fig.~\ref{si-fig:rocksalt_magnetism_phonons_csh}, respectively. In both compounds, phonon magnetic moments appear with a similar orientation along the high-symmetry paths, as also found in NaCl in Fig. 5. Notably, the magnetic moments become substantially larger as the difference in masses increases, going from NaCl to RbF [Fig.~\ref{si-fig:rocksalt_magnetism_phonons_rbf}] and finally CsH [Fig.~\ref{si-fig:rocksalt_magnetism_phonons_csh}].

\clearpage

\section{Other materials}

\subsection*{Computational details}

To demonstrate that chiral surface phonons emerge in a broader class of materials than rocksalt compounds, we also investigate the surface phonons of copper and graphite slabs. These calculations are performed using \texttt{VASP}~\cite{kresseInitioMoleculardynamicsSimulation1994, kresseEfficiencyAbinitioTotal1996, kresseEfficientIterativeSchemes1996} with projector-augmented wave (PAW) pseudopotentials~\cite{kresseUltrasoftPseudopotentialsProjector1999} and the following valence configurations: C (2s$^{2}$2p$^{2}$) and Cu (3d$^{10}$4s$^{1}$). We describe the exchange-correlation interactions using the PBEsol functional~\cite{perdewRestoringDensityGradientExpansion2008} for copper and PBE+D3(BJ)~\cite{perdewGeneralizedGradientApproximation1996, grimmeEffectDampingFunction2011} for graphite. For the bulk calculations, we use a $21 \times 21 \times 21$ $\Gamma$-centered $\mathbf{k}$-mesh and a plane-wave cutoff energy of \SI{600}{\eV} for copper, while graphite is sampled using a $24 \times 24 \times 12$ $\Gamma$-centered $\mathbf{k}$-mesh with a cutoff energy of \SI{800}{\eV}. The bulk structures are optimized by relaxing the lattice parameters of the single-atom primitive unit cell until the total energy and atomic forces converge to within $10^{-6}$ \SI{}{\meV} and $10^{-2}$ \SI{}{\meV\per\angstrom}, respectively. The resulting lattice constant of copper (space group $Fm\bar{3}m$) is $a = \SI{3.57}{\angstrom}$, in close agreement with experimental reports~\cite{braggXLIICrystallineStructure1914, westgrenXXXIXrayAnalysis1925}. Similarly, the optimized lattice parameters of graphite (space group $P6_{3}/mmc$) are $a = \SI{2.47}{\angstrom}$ and $c = \SI{6.71}{\angstrom}$, which are also in good agreement with experimental values~\cite{hasselUeberKristallstrukturGraphits1924}.

The optimized bulk structures are subsequently used to construct the surface slabs. For copper, we consider a 15-layer slab with a (111) surface orientation (space group $P\bar{3}m1$), while for graphite we construct a 10-plane slab with a (0001) surface (space group $P\bar{3}m1$). In both cases, a vacuum region of at least \SI{20}{\angstrom} is introduced along the $z$ direction to minimize interactions between periodic slab images. The Brillouin zone is sampled using $21 \times 21 \times 1$ and $24 \times 24 \times 1$ $\mathbf{k}$-meshes for the copper and graphite slabs, respectively. The slab geometries are then optimized by relaxing the atomic positions while keeping the lattice vectors fixed, using the same convergence criteria as for the bulk structures.

The phonon calculations are carried out with \texttt{phonopy}~\cite{togoFirstprinciplesPhononCalculations2023, togoImplementationStrategiesPhonopy2023}. The interatomic force constants are determined by displacing the atoms by \SI{0.01}{\angstrom}. For copper, we use a $5 \times 5 \times 5$ primitive supercell and a $5 \times 5 \times 5$ $\mathbf{k}$-mesh for the bulk calculations, while the slab calculations employ a $4 \times 4 \times 1$ supercell and a $5 \times 5 \times 1$ $\mathbf{k}$-mesh. For graphite, the bulk calculations use a $6 \times 6 \times 3$ primitive supercell with a $4 \times 4 \times 4$ $\mathbf{k}$-mesh, whereas the slabs are treated using a $4 \times 4 \times 1$ supercell and a $6 \times 6 \times 1$ $\mathbf{k}$-mesh.

\subsection*{Copper}

\subsubsection*{Bulk and surface phonons}

\begin{figure}[htbp]
    \includegraphics{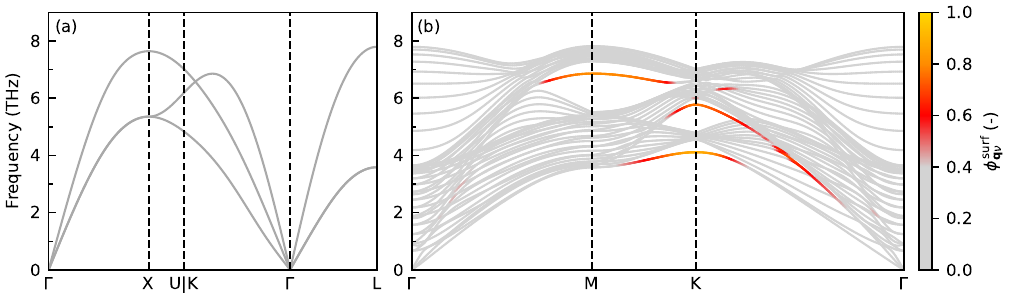}
    \caption{Phonons in copper. (a) Bulk phonon dispersion. (b) Slab phonon dispersion of a 15-layer (111)-oriented slab. The phonon bands are colored with the surface localization $\phi^{\mathrm{surf}}_{\mathbf{q} \nu}$.}
    \label{si-fig:copper_phonons}
\end{figure}

The bulk and slab phonon dispersion of copper are shown in Fig.~\ref{si-fig:copper_phonons}. Figure~\ref{si-fig:copper_phonons}(a) shows the phonon dispersion of bulk copper. Unlike materials with a rocksalt structure, bulk copper exhibits only three acoustic phonon branches, as a direct consequence of its single-atom primitive unit cell. Creating a 15-layer (111)-oriented slab, increases the total number of phonon bands to 45 [Fig.~\ref{si-fig:copper_phonons}(b)]. Several of these bands are strongly localized at the surface and can therefore be identified as surface phonons. We note that the calculated surface phonon bands are in good agreement with previous studies of surface phonons in copper~\cite{chisLargeSurfaceCharge2008}.

\subsubsection*{Chiral phonons}

\begin{figure}[htbp]
    \includegraphics{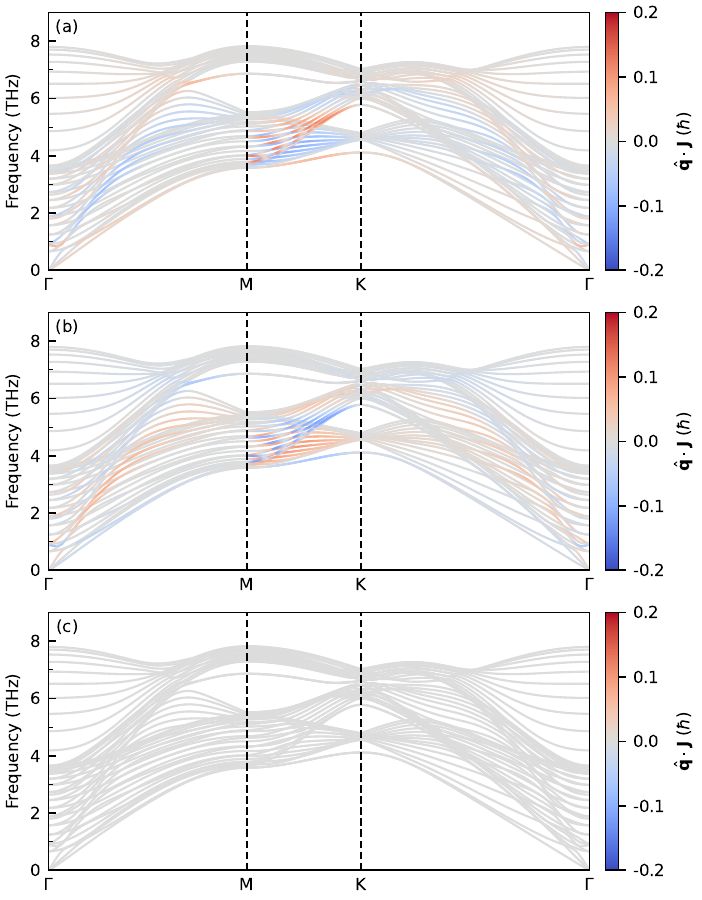}
    \caption{Chiral phonons in 15-layer (111)-oriented copper slabs. Slab phonon dispersion, with the phonon bands colored with the phonon chirality $\hat{\mathbf{q}} \cdot \mathbf{J}_{\mathbf{q} \nu}$ of (a) the top, (b) the bottom, and (c) both surface layers combined.}
    \label{si-fig:copper_chiral_phonons_combined}
\end{figure}

Next, we calculate the chirality of Cu(111) surface phonons. The phonon dispersion, with the color-coding representing the phonon chirality of the top of the slab, is shown in Fig.~\ref{si-fig:copper_chiral_phonons_combined}(a). We find that the surface phonon modes along the $\mathrm{\Gamma - M}$ at and $\mathrm{M - K}$ paths acquire a finite phonon chirality, at around \SI{6.5}{\THz} and \SI{3.7}{\THz}, respectively. Importantly, the bottom surface layers of the slab exhibit exactly the opposite phonon chirality [Fig.~\ref{si-fig:copper_chiral_phonons_combined}(b)], resulting in a net zero phonon chirality when considering the entire slab [Fig.~\ref{si-fig:copper_chiral_phonons_combined}(c)]. This demonstrates that chiral surface phonons are not restricted to insulating rocksalt structures.

\subsection*{Graphite}

\subsubsection*{Bulk and surface phonons}

\begin{figure}[htbp]
    \includegraphics{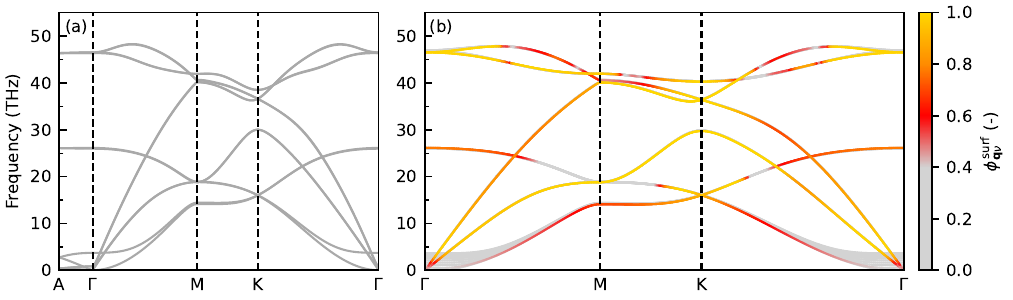}
    \caption{Phonons in graphite. (a) Bulk phonon dispersion. (b) Slab phonon dispersion of a 10-layer (0001)-oriented slab. The phonon bands colored with the surface localization $\phi^{\mathrm{surf}}_{\mathbf{q} \nu}$.}
    \label{si-fig:graphite_phonons}
\end{figure}

For graphite, the bulk and slab phonon dispersions are shown in Fig.~\ref{si-fig:graphite_phonons}. In the bulk phonon dispersion, shown in Fig.~\ref{si-fig:graphite_phonons}(a), there are six phonon bands, which stem from the two-atom unit cell of graphite. In contrast, the 10-layer (0001)-oriented slab contains 20 atoms, and therefore the slab dispersion in Fig.~\ref{si-fig:graphite_phonons}(b) contains a total of 60 phonon bands. As a result of the weak interlayer coupling between the graphite sheets, the surface graphite layers are rather similar to the bulk layers, resulting in a minimal decrease in energy for the surface modes at high energies compared to their energies in the bulk. In agreement with earlier work~\cite{derouffignacLatticeDynamicsGraphite1981}, we only observe a substantial change in energy for the low-frequency bands close to $\Gamma$.

\subsubsection*{Chiral phonons}

\begin{figure}[htbp]
    \includegraphics{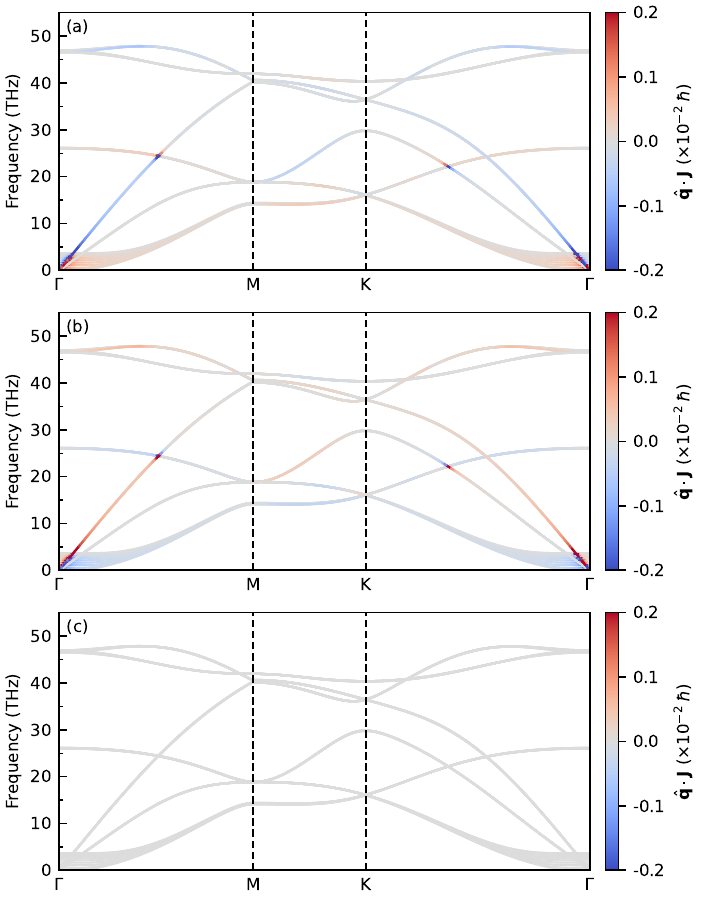}
    \caption{Chiral phonons in 10-layer (0001)-oriented graphite slabs. Slab phonon dispersion, with the phonon bands colored with the phonon chirality $\hat{\mathbf{q}} \cdot \mathbf{J}_{\mathbf{q} \nu}$ of (a) the top, (b) the bottom, and (c) both surface layers combined.}
    \label{si-fig:graphite_chiral_phonons_combined}
\end{figure}

Next, we examine whether the phonons of the (0001)-oriented graphite surface exhibit chirality. As for the other materials, we analyze the phonon chirality of the surface layers of the graphite slab. The resulting phonon dispersion, with the color-coding indicating the phonon chirality of the surface layers, is shown in Fig.~\ref{si-fig:graphite_chiral_phonons_combined}. As with the rocksalt and copper slabs, we find that the surface phonons of graphite are also chiral, i.e. $\hat{\mathbf{q}} \cdot \mathbf{J} \neq 0$. This demonstrates that chiral surface phonons can also emerge in layered compounds such as graphite. However, in comparison to the other materials, the surface phonons of graphite have a chirality that is 1 to 2 order of magnitude smaller. This reduced chirality originates from the weak interlayer coupling between the graphite layers. Consequently, the components of the force constant matrix that couple in-plane and out-of-plane displacements are already small in the bulk crystal. Reducing these components to zero at the surface therefore introduces only a small phonon chirality.

\clearpage

\bibliography{references}